# Upper critical field, pressure-dependent superconductivity and electronic anisotropy of $Sm_4Fe_2As_2Te_{1-x}O_{4-y}F_y$

A Pisoni, S Katrych, P Szirmai, B Náfrádi, R Gaál, J Karpinski and L Forró

Institute of Condensed Matter Physics, EPFL, CH-1015 Lausanne, Switzerland

E-mail: andrea.pisoni@epfl.ch

## Abstract

We present a detailed study of the electrical transport properties of a recently discovered iron-based superconductor: $Sm_4Fe_2As_2Te_{0.72}O_{2.8}F_{1.2}$. We followed the temperature dependence of the upper critical field by resistivity measurement of single crystals in magnetic fields up to 16 T, oriented along the two main crystallographic directions. This material exhibits a zero-temperature upper critical field of 90 T and 65 T parallel and perpendicular to the $Fe_2As_2$ planes, respectively. An unprecedented superconducting magnetic anisotropy $\gamma_H = H_{c2}^{ab}/H_{c2}^c \approx 14$ is observed near $T_c$, and it decreases at lower temperatures as expected in multiband superconductors. Direct measurement of the electronic anisotropy was performed on microfabricated samples, showing a value of $\rho_c/\rho_{ab}(300K) \approx 5$ that raises up to 19 near $T_c$. Finally, we have studied the pressure and temperature dependence of the in-plane resistivity. The critical temperature decreases linearly upon application of hydrostatic pressure (up to 2 GPa) similarly to overdoped cuprate superconductors. The resistivity shows saturation at high temperatures, suggesting that the material approaches the Mott-Ioffe-Regel limit for metallic conduction. Indeed, we have successfully modelled the resistivity in the normal state with a parallel resistor model that is widely accepted for this state. All the measured quantities suggest strong pressure dependence of the density of states.

## 1. Introduction

We recently succeeded in synthesizing a new class of pnictide superconductors: $L_4Fe_2As_2Te_{1-x}O_{1-y}F_y$ ($L$=Pr, Sm, Gd) [1, 2], which we have labelled "42214" following the convention described in ref. 1 . These compounds present critical temperatures ($T_c$) up to 46 K upon substitution of the rare earth element and optimum fluorine doping. Their tetragonal crystal structure presents an alternation of $Fe_2As_2$ conducting layers and $L_2O_2$ spacing layers separated along the $c$ direction by Te atoms. The lattice constants $a$ and $b$ are comparable to those found in the "1111" structure but the $c$ axis parameter is ~3.5 times larger because of the oxide layer [3]. The presence of tellurium vacancies is thought to be responsible for superconductivity occurring at 25 K in $L_4Fe_2As_2Te_{1-x}O_4$ samples. Here we present a systematic and comprehensive study of the electronic and magnetic properties of $Sm_4Fe_2As_2Te_{1-x}O_{1-y}F_y$ which displays $T_c$=40 K upon optimum electron doping by fluorine substitution [2]. We report the upper critical field, pressure-dependent resistivity and direct measurement of the electronic anisotropy of $Sm_4Fe_2As_2Te_{0.72}O_{2.8}F_{1.2}$.

## 2. Experimental details

Single crystals with nominal composition $Sm_4Fe_2As_2Te_{0.72}O_{2.8}F_{1.2}$ were grown at high temperature and at high pressure employing a cubic-anvil system. Details regarding the crystals growth and their characterization are described elsewhere [1, 2]. To measure electrical resistivity ($\rho$) a standard four-point configuration was adopted. Due to the small sample dimensions (< 100 µm) platinum leads were deposited on the sample surfaces by focused ion beam (FIB) technique. Magnetic field was applied both parallel and perpendicular to the crystallographic *c*-axis. Resistivity was recorded as a function of temperature down to 1.7 K in different magnetic fields up to 16 T. The temperature dependence of $\rho$ in hydrostatic pressures up to 2 GPa was measured in a piston-cylinder cell employing Daphne oil 7373 as pressure transmitting medium. The pressure was determined by the superconducting transition temperature of a lead pressure gauge. To measure the electronic anisotropy ($\rho_c/\rho_{ab}$), FIB was used to shape a lamella extracted from the single crystal. This has allowed simultaneous 4 point measurement of the two resistivities on the same sample.

## 3. Results and discussion

### 3.1 Upper critical field

In order to extract the upper critical field ($H_{c2}$) of $Sm_4Fe_2As_2Te_{0.72}O_{2.8}F_{1.2}$, resistivity measurements were performed as a function of temperature at different constant magnetic fields up to 16 T. The results for magnetic field oriented perpendicular to the $Fe_2As_2$ layers ($H \parallel c$) and along them ($H \parallel ab$) are reported in figure 1(a) and (b), respectively. In zero-field the superconducting temperature for our crystals is $T_c \sim 35$ K. With increasing magnetic field $T_c$ decreases showing the usual suppression of superconductivity. As already observed in other layered superconductors, this suppression is more pronounced when $H \parallel c$ than when $H \parallel ab$ [4]. For $H \parallel c$ the resistivity shows a small up-turn that increases with magnetic field. This could be ascribed to possible disorder or fluorine doping inhomogeneity inside the sample. The width of the superconducting transition monotonically increases with magnetic field parallel to *ab* planes while for $H \parallel c$ an anomalous sharpening of the superconducting transition is observed upon increasing field. In order to confirm this anomaly we measured $H_{c2} \parallel c$ for another sample shaped by FIB in a rectangular bar only 10 µm long and $5 \times 5$ µm in cross-section. The results (see supplemental material) show the same characteristics. A similar phenomenon was already reported in other unconventional superconductors [5] and in $NdFeAsO_{0.7}F_{0.3}$ for magnetic field above 25 T [6].

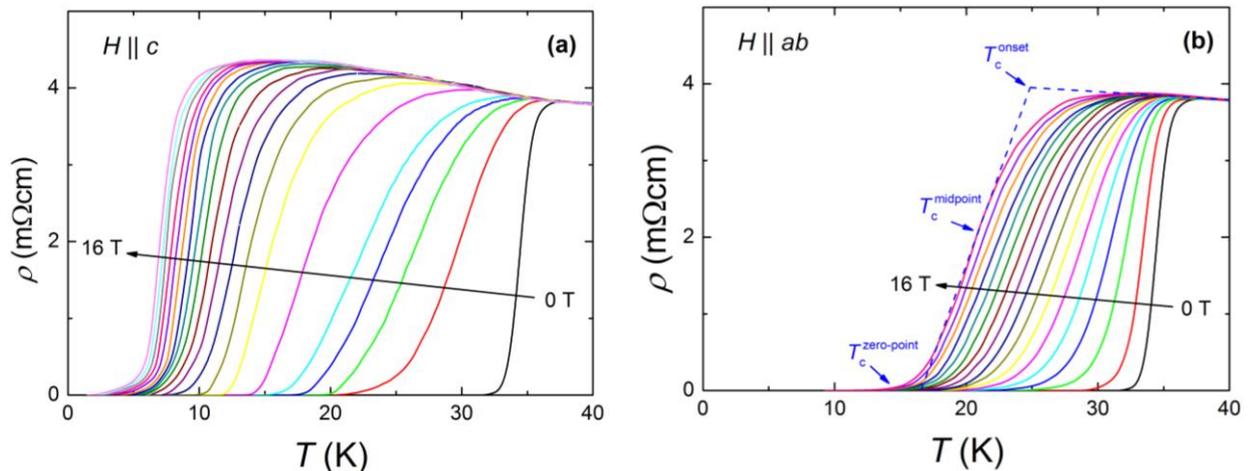

**Figure 1.** Temperature and magnetic field dependence of resistivity of $Sm_4Fe_2As_2Te_{0.72}O_{2.8}F_{1.2}$ single crystal measured with fields applied (a) perpendicular to the $Fe_2As_2$ layers ($H \parallel c$) and (b) parallel to them ($H \parallel ab$), the dashed blue line represent the criteria used to evaluate the three different critical temperatures as stated in the text. The values of the magnetic field are 0, 0.25, 0.5, 1, 2, 3, 4, 5, 6, 7, 8, 9, 10, 11, 12, 13, 14, 15 and 16 T following the arrow in both panels.

Figure 2 presents the upper critical magnetic field of $Sm_4Fe_2As_2Te_{0.72}O_{2.8}F_{1.2}$, from resistivity measurements, as a function of temperature. In our analysis we define the onset $T_c$ at the crossing point of two extrapolated lines: one drawn through the resistivity curve in the normal state just above $T_c$, and the other through the steepest part of the resistivity curve in the superconducting state. The midpoint $T_c$ is determined as the temperature at which the resistivity is 50% of its value at the onset $T_c$. The zero-point $T_c$ is defined at zero-resistivity point (i.e., when the measured $\rho(T)$ drops below our experimental sensitivity). The graphical representation of these three different critical temperatures is provided in figure 1(b). From these three temperatures, we calculate the transition width as the difference between the onset and zero-point $T_c$. The upper critical field ($H_{c2}$) is defined at the midpoint $T_c$ and it is shown as a function of temperature for $H \parallel c$ and $H \parallel ab$ (filled dots). The empty dots in figure 2 represent the values of $H_{zero\ res.}$ defined at the zero-resistivity point. The temperature dependences of the in-plane ($H_{c2}^{ab}$) and the out-of-plane ($H_{c2}^{c}$) upper critical field show a non-linear trend with a convex form more pronounced for $H \parallel c$. This behaviour was already reported in some "1111" superconducting compounds [7] and it was interpreted as an intrinsic feature of pnictide superconductors, explained by a two band model [6]. Because of the positive second derivative, $H_{c2}(T)$ cannot be explained by the standard Werthamer-Helfand-Hohenberg (WHH) model [8] that is strictly valid for single band conventional superconductors in weak-coupling regime and that is found to underestimate $H_{c2}(0)$ considerably in unconventional superconductors [6, 9-11]. We therefore decided to extract the zero-temperature values $H_{c2}^{c}(0)$ and $H_{c2}^{ab}(0)$ using a theoretical approach usually employed for cuprate superconductors [12-16] and other unconventional superconductors [17, 18] with a very similar $H_{c2}(T)$ curve. Following the example of Zhang et al. [13], we fitted the critical field values in figure 2 with

$$H_{c2}(T) = A + Be^{(-T/C)} \qquad (1)$$

where A, B and C are constant parameters. This purely empirical functional form, besides describing the data points sufficiently well, has the advantage of yielding a finite $H_{c2}(0)$. The resulting fits are presented in blue dashed lines in figure 2. As it can be seen, equation (1) gives a remarkably good description of the experimental results. From the above fitting we obtained $\mu_0 H_{c2}^{ab}(0) = \Phi_0/(2\pi\xi_{ab}\xi_c) = 90\,\text{T}$, where $\xi_{ab}$ is the in-plane coherence length, $\xi_c$ the out-of-plane coherence length and $\Phi_0$ is the magnetic flux quantum. For $H \parallel c$ we deduced $\mu_0 H_{c2}^{c}(0) = \Phi_0/(2\pi\xi_{ab}^2) = 65\,\text{T}$, which corresponds to $\xi_{ab} = 22.5\,\text{Å}$, therefore implying $\xi_c = 16.3\,\text{Å}$. The magnetic field anisotropy ($\gamma_H$) was extracted using both the midpoint criterion (filled blue dots) and the zero-resistivity criterion (empty blue dots) and it is presented in the inset of figure 2. As it can be observed the two criteria give $\gamma_H$ values that follow a common and continuous trend. The solid black line, in the inset of figure 2, represents the values of $\gamma_H$ calculated from the fits of the $H_{c2}(T)$ curves, as described above. The magnetic field anisotropy shows a maximum at $T \sim 0.75 * T_c = 25$ K, as reported in other pnictide superconductors [19, 20] and which is in agreement with some theoretical model [21] for multiband superconductors. The value of the maximum, $\gamma_H = H_{c2}^{ab}/H_{c2}^{c} \sim 14$, is the highest reported so far among iron-based superconductors [6, 11, 20, 22] and it reflects the remarkable structural anisotropy of this oxypnictide compound. With lowering temperature $\gamma_H$ decreases to a nearly isotropic system as reported for other Fe-based superconductors [23, 24].

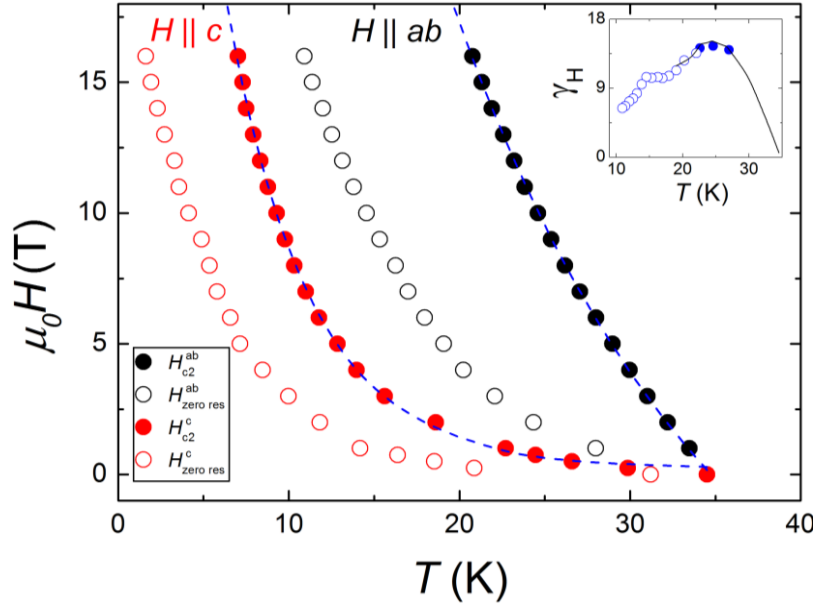

**Figure 2.** Magnetic field-temperature phase diagram from resistivity measurement with $H \parallel ab$ (black points) and $H \parallel c$ (red points). Filled dots correspond to values of the upper critical field ($H_{c2}$) estimated from the midpoint of the resistive transitions while empty dots represent the fields evaluated at the zero-resistivity point. Blue dashed lines are the fits (as explained in the text) used to extract the values of $H_{c2}^{ab}(0)$ and $H_{c2}^{c}(0)$. The inset shows the magnetic field anisotropy ($\gamma_H$) extracted using the midpoint criterion (filled blue dots) and the zero-resistivity criterion (empty blue dots) as a function of temperature near $T_c$. The solid black line in the inset represents the value of $\gamma_H$ calculated from the fitted function as described in the text.

*3.2 Pressure effect*

Figure 3 shows $\rho(T)$ at different hydrostatic pressures up to 2 GPa. The room temperature value of $\rho$ decreases from $\rho(300K) = 7.55$ mΩcm at 0 GPa to $\rho(300K) = 6$ mΩcm at 2 GPa. The resistivity in the normal state presents a non-linear behaviour that flattens out approaching a limiting value ($\rho_{sat}$) as the temperature increases. Such a saturation phenomenon was already observed and extensively studied in many A15 superconductors [25, 26], Chevrel phases [27] and $V_xSi_y$ [28] alloys. It was interpreted as evidence of the Mott-Ioffe-Regel (MIR) criterion [29, 30], which states that the electron mean free path ($l$) cannot be shorter than the interatomic distance ($a$). In order to justify the analysis based on the MIR criterion we fitted our experimental data employing the "parallel-resistor model" [31, 32], a phenomenological model in which the limiting resistivity is reached when $l \approx a$:

$$\frac{1}{\rho(T)} = \frac{1}{\rho_{id}(T)} + \frac{1}{\rho_{sat}} \qquad (2)$$

with

$$\rho_{id}(T) = \rho_{id}(0) + AT \qquad (3)$$

where $\rho_{id}(0)$ is the ideal temperature independent factor due to scattering by impurities and defects, while the linear term in temperature is associated with the phonon-scattering. In metals where $l \gg a$, like in copper ($l_{Cu} \approx 100$ Å) [33], $\rho(T) \approx \rho_{id}(T)$. In figure 3 the blue solid lines are the least squares fits to the data using equation (2) and (3), where $\rho_{id}(0)$, $A$ and $\rho_{sat}$ are allowed to vary. As it can be seen, the agreement between the model and the experimental data is very good. The values of $\rho_{id}(0)$, $A$ and $\rho_{sat}$ are reported in table 1 for each pressure. The values of $\rho_{sat}$ appear to be nearly independent of the applied pressure, while the values of $A$ and $\rho_{id}(0)$ decrease monotonously as the pressure increases. In A15 compounds [30, 32] $\rho_{sat}$ is experimentally found between 100 and 300 µΩcm, in good agreement with the MIR estimation. These values are an order of magnitude lower than the one we observe in $Sm_4Fe_2As_2Te_{0.72}O_{2.8}F_{1.2}$. However, previous reports in both $La_{2-x}Sr_xCuO_4$ [33] (for small x) and $Bi_2Sr_2Ca_{1-x}Y_xCu_2O_{8+y}$ [34] showed saturation to resistivity values in the mΩcm range at room temperature, similar to our case. Gunnarsson *et al.* [33] demonstrated that the semiclassical approach used for the MIR criterion breaks down when one deals with strongly correlated multiband systems where the interband transitions are important. They employed the $t-J$ model to show that resistivity can saturate at much higher values than predicted by the MIR criterion [33]. In materials with a quasi-two dimensional band with cylindrical Fermi surface the MIR criterion can be easily checked using the relation $k_F l \simeq 0.258 \left( \dfrac{\delta}{\rho} \right)$, where $k_F$ is the Fermi wavevector, $\rho$ the electrical resistivity expressed in mΩcm and $\delta$ is the distance between the conducting layers expressed in Å [35]. The MIR criterion is satisfied if $k_F l > 1$ [33]. In the case of $Sm_4Fe_2As_2Te_{0.72}O_{2.8}F_{1.2}$, $\delta = 11.924$ Å, resulting in $k_F l < 1$. The fact that saturation occurs at high resistivity values, in agreement with what already observed in other Fe-based superconductors [36], could be ascribed to the multiband nature of these superconductors.

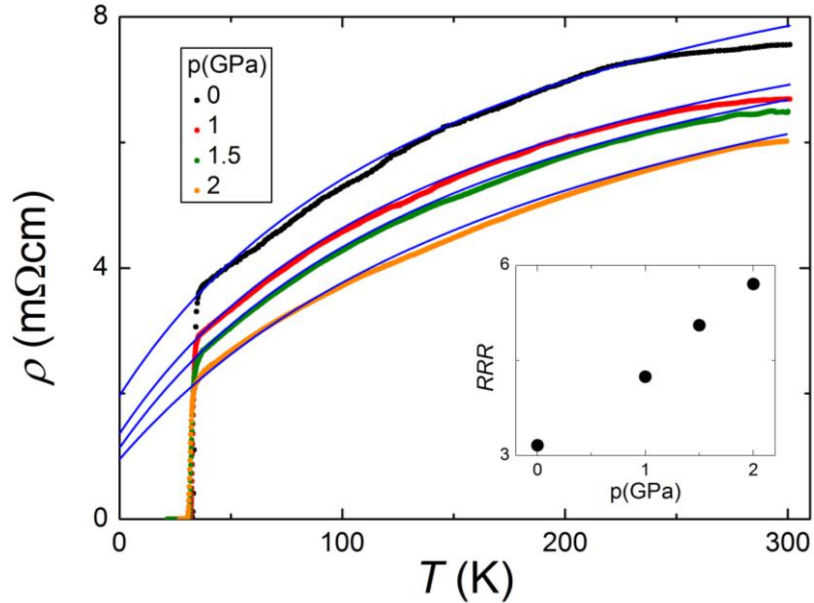

**Figure 3.** Electrical resistivity versus temperature at different pressures up to 2 GPa. The inset shows the pressure dependent residual resistance ratio. The solid blue lines represent fits to the parallel-resistor model as explained in the text.

**Table 1.** Fitting parameters for the parallel-resistor model as explained in the text.

| p (GPa) | $\rho_{id}(0)$ (mΩcm) | $A$ (mΩcmK$^{-1}$) | $\rho_{sat}$ (mΩcm) |
|---|---|---|---|
| 0 | 2.388 | 0.0821 | 11.073 |
| 1 | 1.577 | 0.0755 | 9.879 |
| 1.5 | 1.282 | 0.0643 | 9.888 |
| 2 | 1.054 | 0.0503 | 9.893 |

The residual resistance ratio $RRR = \rho(300\text{K})/\rho(0\text{K})$, where $\rho(0\text{K}) = \rho_{id}(0)$ is value of resistivity extrapolated at zero Kelvin using the parallel-resistor model, increases from 3.2 to 5.7 with pressure (inset figure 4). This variation is in large part due to the change of the zero temperature term. In figure 4 the variation of $T_c$ with pressure is presented. $T_c$ decreases almost linearly at a rate of 1 K/GPa keeping a constant superconducting transition width of 2 K up to 2 GPa. The linear-like decrease of the critical temperature with pressure in Fe-based superconductors has been interpreted earlier as an indication of overdoping similarly to the case of cuprates [37, 38]. Conventional BCS theory predicts a decrease of $T_c$ at high pressure according to $T_c \sim \Theta_D \cdot \exp[-1/D(E_F)U]$ where $\Theta_D$ is the Debye temperature, $D(E_F)$ the electronic density of states at the Fermi energy and $U$ is the potential of electron-phonon interaction. Since in solids $\Theta_D$ increases with pressure (phonon hardening) [39], the observed decrease of $T_c$ must be caused by a decrease of $D(E_F)$. In fact the density of states is expected to decrease under pressure due to the band broadening induced by pressure [40]. However, the improvement of $RRR$ with pressure in $Sm_4Fe_2As_2Te_{0.72}O_{2.8}F_{1.2}$ demonstrates that the pressure effect is more complex than a simple change in carrier concentration.

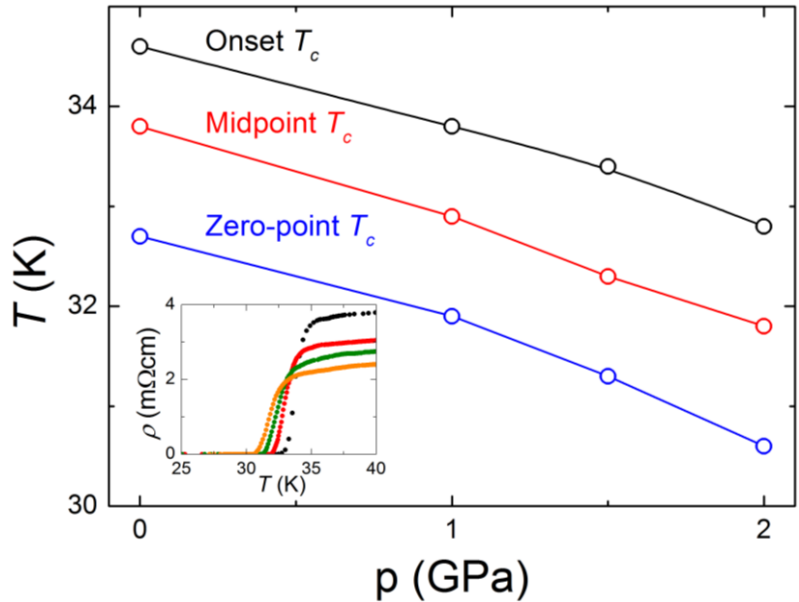

**Figure 4.** Critical temperatures evaluated with different criteria (see text) as a function of pressure. Continuous lines are guide to the eye. The inset shows the behavior of $\rho$ near the transition.

### 3.3 Electronic anisotropy

The measurement of electrical anisotropy required microfabrication techniques due to the small sizes of the available samples. Inspired by the work of Moll *et al.* [41], we employed Focused Ion Beam to cut a lamella out from the $Sm_4Fe_2As_2Te_{0.72}O_{2.8}F_{1.2}$ crystal. The lamella was placed on an insulating substrate ($SiO_2$) and ion-assisted platinum deposition was used for placing electrical leads. The voltage probes were 1-2 µm apart along the *c*-axis direction and 12 µm along the *ab* plane. The lamella was then shaped to the structure shown in figure 5. In this configuration we achieved a direct and simultaneous measurement of the resistivity along *c*-axis ($\rho_c$) and *ab* plane ($\rho_{ab}$) on the same sample and under the same conditions.

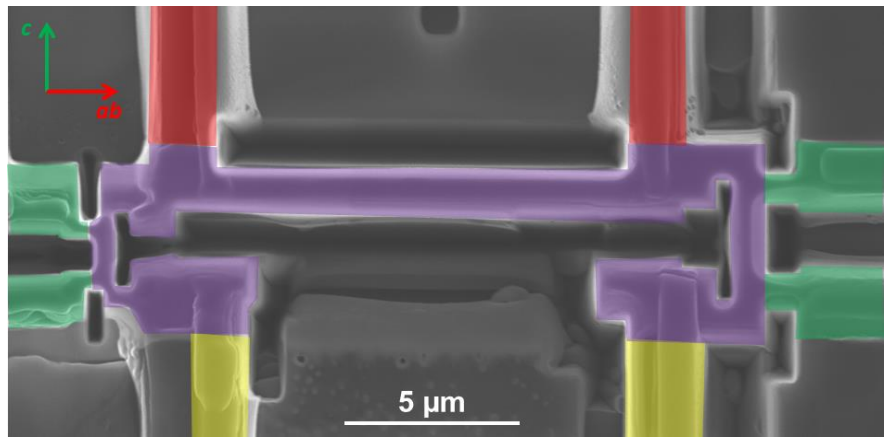

**Figure 5**. Scanning electron micrograph of the FIB cut lamella extracted from $Sm_4Fe_2As_2Te_{0.72}O_{2.8}F_{1.2}$ single crystal. The U-shaped crystal (violet) is electrically connected with platinum leads from which voltage is measured along *c*-axis (green leads) and *ab* plane (red leads) when a d.c.-current is applied through the yellow leads.

Figure 6 shows the temperature dependence of $\rho_c$ (green) and $\rho_{ab}$ (red). The micro-structured lamella presents a slightly lower mid-point $T_c \sim 32$ K than measured for the single crystal from which it was extracted ($T_c \sim 34$ K). This could be due to ion damage induced by FIB. $\rho_c$ increases from 27 mΩcm at 300 K to 74 mΩcm at $T_c$ displaying a tendency to localization at low temperatures. A very similar observation was reported for another iron-based superconductor, $(V_2Sr_4O_6)Fe_2As_2$, that displays an intrinsic Josephson junction behaviour [42]. However, it's unlikely that intrinsic Josephson junction effect could occur in $Sm_4Fe_2As_2Te_{1-x}O_{4-y}F_y$ because of the short $Fe_2As_2$ inter-layer distance ($\delta$) compared to the out-of-plane coherence length ($\delta/\xi_c \square 0.7$). The *ab* plane resistivity decreases with lowering temperature and a mild upturn is observed just above $T_c$, exactly as observed in the single crystal from which the lamella was extracted [2]. This upturn could come from slight admixture of the out-of-plane resistivity which gives a different appearance of $\rho_{ab}$ than in figure 2. The absolute value of $\rho_{ab}$ is in very good agreement with the resistivity value observed in the single crystal [2], confirming the dominant electronic transport along the *ab* planes. A previous work on $SmFeAsO_{0.7}F_{0.25}$ shows the same trend for $\rho_c$ and $\rho_{ab}$ [41]. The resistivity anisotropy ($\rho_c/\rho_{ab}$) is plotted as function of temperature in the inset of figure 6. As observed in other quasi-two dimensional superconductors $\rho_c/\rho_{ab}$ increases on cooling indicating that charge carriers are more and more confined in the FeAs layers when the temperature is decreases. A previous work on SmFeAsO$_{0.7}$F$_{0.25}$ shows the same trend for $\rho_c$ and $\rho_{ab}$ [41]. In SmFeAsO$_{0.7}$F$_{0.25}$ the room temperature resistivity ratio is $\rho_c/\rho_{ab} = 2$ and it increases to 10 at $T_c$ while in our case $\rho_c/\rho_{ab}(300K) = 5$ and $\rho_c/\rho_{ab}(T_c) = 19$. The value of electronic anisotropy in Sm$_4$Fe$_2$As$_2$Te$_{0.72}$O$_{2.8}$F$_{1.2}$ is low when compared to that observed in cuprate superconductors and in some organic compounds [43-45]. This could be due to the tellurium vacancies, naturally present in the spacing layers of this compound, which could act as a tunnelling bridge between adjacent Fe$_2$As$_2$ planes. The fluctuation-tunnelling model, developed by Sheng [46], can describe well the conduction of charge carriers along the $c$ direction. According to this model the resistivity can be expressed as

$$\rho(T) = \rho_0 \exp\left(\frac{T_0}{T_1 + T}\right) \tag{4}$$

where

$$T_0 = \frac{16\varepsilon_0 \hbar A V_0^{3/2}}{\pi e^2 k_B w^2 \sqrt{2m}} \tag{5}$$

$$T_1 = \frac{8\varepsilon_0 A V_0^2}{e^2 k_B w} \tag{6}$$

where $k_B$ is the Boltzmann constant, $\varepsilon_0$ is the vacuum permittivity, $w$ is the tunnelling junction width, $m$ is the electron effective mass, $A$ is the tunnelling junction area and $V_0$ is the tunnelling height barrier. The temperature dependence of $\rho_c$ can be fitted very well to equation (4) (blue line in figure 6), using $\rho_0 = 16.4$ m$\Omega$cm, $T_0 = 220$ K and $T_1 = 115$ K. Assuming an electron effective mass equal to the free electron mass and a value of $w$ equal to the Fe$_2$As$_2$ inter-layer distance, it's possible to estimate $A$ and $V_0$ (from equations (5) and (6)) using the fitting parameters mentioned above. Our results give $A = 3.8 \times 10^{-3}$ μm$^2$ and a tunnelling energy barrier of 3 meV.

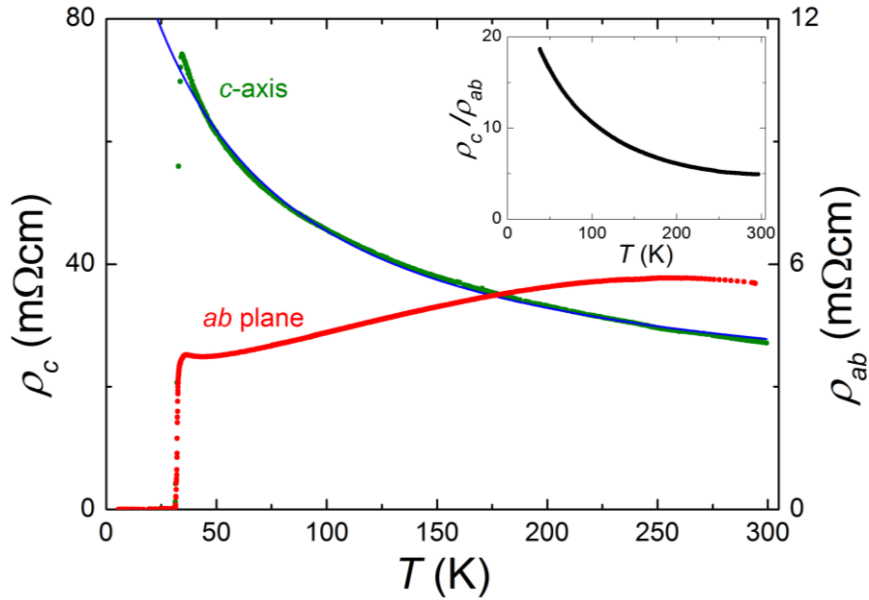

**Figure 6.** Temperature dependence of resistivity measured along different crystallographic directions. The red curve represents the resistivity measured along the $ab$ plane and it refers to the scale of the right, while the green

curve represents the resistivity measured along the *c* direction and it refers to the scale on the left. The inset shows the electrical anisotropy as a function of temperature.

## 4. Conclusions

In conclusion we performed a systematic and detailed investigation of the superconducting properties of $Sm_4Fe_2As_2Te_{01-x}O_{1-y}F_y$. Resistivity measurements in magnetic fields up to 16 T reveal upper critical fields of $\mu_0 H_{c2}^c(0) = 65\,\text{T}$ and $\mu_0 H_{c2}^{ab}(0) = 90\,\text{T}$ parallel and perpendicular to the crystallographic *c*-axis direction. We observe unprecedented magnetic field anisotropy near $T_c$, $\gamma_H = 14$, more than twice higher than in $SmFeAsO_{1-y}F_y$. With decreasing temperature, $\gamma_H \to 1$ confirming the multiband, unconventional superconducting character of $Sm_4Fe_2As_2Te_{01-x}O_{1-y}F_y$. The measurement of electrical properties under hydrostatic pressures up to 2 GPa demonstrates a linear decrease of $T_c$ with increasing pressure, indicating a sample in the overdoped regime. The normal state resistivity shows saturation at high temperatures, indicative of approaching the Mott-Ioffe-Regel limit. Indeed, $\rho(T)$ is well fitted by the widely accepted parallel resistor model. The high temperature limiting value of the resistivity is very high (~10 mΩcm), but still compatible with metallic conduction if the strongly anisotropic structure of the material is taken into account. The variation of the fitting parameters with pressure, in particular that of the impurity scattering suggests strong changes in the normal carrier density with pressure. Direct and reliable measurement of the electronic anisotropy was possible by FIB micromachining of lamella extracted from $Sm_4Fe_2As_2Te_{0.72}O_{2.8}F_{1.2}$ single crystal. The electronic anisotropy displays a very similar behaviour to that previously reported for $SmFeAsO_{1-y}F_y$ except for the considerably larger absolute value (more than twice). The temperature dependence of the resistivity along the *c* direction is interpreted by a fluctuation-tunnelling model where charges carriers tunnel between different tellurium vacancies naturally present in the crystal structure of this compound. The large values of the *c* lattice parameter, of the critical field anisotropy and of the electrical anisotropy place $Sm_4Fe_2As_2Te_{1-x}O_{4-y}F_y$ among the most anisotropic known pnictide superconductors

**Acknowledgements**

This work was supported by the Swiss National Science Foundation (Projects No. 140760, No. 138053 and No. 156012).